\documentstyle[preprint,aps]{revtex}

\tighten
\begin{document}
\preprint{\parbox[t]{90mm}{
Preprint Number: \parbox[t]{50mm}{ANL-PHY-7711-94}}}
\title{Gauge covariant fermion propagator in\\  quenched, chirally-symmetric
quantum electrodynamics}

\author{Zhihua Dong\footnotemark[2], Herman J. Munczek\footnotemark[2] and
Craig D. Roberts\footnotemark[1]\vspace*{2mm} }

\address{\footnotemark[2]Department of Physics and Astronomy,
University of Kansas, \\
Lawrence, Kansas 66045-2151, USA\vspace*{2mm}\\
\footnotemark[1]Physics Division, Argonne National Laboratory,
Argonne, Illinois 60439-4843, USA}

%
\maketitle
%
\begin{abstract}
We discuss the chirally symmetric solution of the massless, quenched,
Dyson-Schwinger equation for the fermion propagator in three and four
dimensions.  The solutions are manifestly gauge covariant.  We consider a
gauge covariance constraint on the fermion--gauge-boson vertex, which
motivates a vertex Ansatz that both satisfies the Ward identity when the
fermion self-mass is zero and ensures gauge covariance of the fermion
propagator.
\end{abstract}
%
\subsection{Introduction}
The Dyson-Schwinger equations (DSEs) are a valuable nonperturbative tool for
studying field theories; recent reviews of this approach can be found in
Refs.~\cite{H91,R93,RW94}.  While this approach has limitations, which are
being addressed, it greatly facilitates the development of models which
bridge the gap between short-distance, perturbative QCD and the extensive
amount of low- and intermediate-energy phenomenology in a single, covariant
framework.  In addition, as the approach continues to be developed, there are
obvious points where cross-fertilisation with lattice studies will become
valuable.  One such example is the study of the phenomenological implications
of gluon propagators obtained in lattice simulations\cite{HRW94}.

In recent years a good deal of progress has been made in addressing the
limitations of the DSE approach in the study of Abelian gauge
theories. Calculations are now such that direct comparison can be made
between quantities calculated in the DSE approach and those calculated in
lattice simulations.  One illustrative example is the gauge invariant fermion
condensate, \mbox{$\langle\bar\psi \psi\rangle$}.  This has been calculated
in three-dimensional QED and the agreement with reported lattice
results\cite{DKK90} is very good\cite{RW94,B93}.  This progress has been made
possible by a realisation of the importance of the fermion--gauge-boson
vertex.

The fermion--gauge-boson vertex satisfies a DSE. This equation involves the
kernel of the fermion-antifermion Bethe-Salpeter equation, which cannot be
expressed in a closed form; i.e., its skeleton expansion has infinitely many
terms.  Given this, research has concentrated on placing physically
reasonable constraints on the form of the vertex, constructing simple
Ans\"atze that embody them, employing a given Ansatz in the DSE for the
fermion propagator, and studying the properties of the solution.  This has
lead to an understanding of the role of the vertex in ensuring multiplicative
renormalisability\cite{CP90} and gauge covariance\cite{BR93}, within the
class of linear, covariant gauges.

Recently a gauge covariance constraint on a class of vertices, applicable to
chirally-symmetric, quenched QED, was proposed\cite{BR93}.  One
characteristic of this class of vertices is that the DSE admits the free
propagator solution in Landau gauge.  This constraint was used in a critical
analysis of three often used vertex Ans\"atze.  Of these three, that of
Ref.~\cite{CP90} was the only one not eliminated.  [Herein chirally symmetric
means that that the fermion bare mass is zero and there is no dynamical mass
generation and quenched means that the vacuum polarisation is neglected.]
This Ansatz has been used in studies of three- and four-dimensional
QED\cite{BR93} and in phenomenological studies of QCD\cite{HRW94}.  The
fermion-DSE in four-dimensional, chirally symmetric, quenched QED was first
studied using this Ansatz in Ref.~\cite{CP91}, however, the explicit form of
the solution obtained therein is not gauge covariant, inconsistent with the
expectations of Ref.~\cite{BR93}.

Herein we are interested in analysing, and making explicit, the restrictions
on the fermion--gauge-boson vertex imposed by the gauge covariance constraint
proposed in Ref.~\cite{BR93}.  To this end, we solve the fermion-DSE in
three- and four-dimensional QED and obtain manifestly gauge covariant
solutions in both cases.  In the process we show how the constraint is
crucial to this outcome and identify an error in Ref.~\cite{CP91} that leads
to the disagreement with Ref.~\cite{BR93}: it arises because of an
inappropriate regularisation procedure.  We show that, although the
gauge-covariance constraint is satisfied by the chirally-symmetric limit of
the Ansatz of Ref.~\cite{CP90}, this vertex violates the Ward identity and is
hence unsuitable in this application.  We use the constraint to construct an
Ansatz that overcomes this defect.  Our study shows that the gauge-covariance
constraint allows a large class of vertex Ans\"atze, of which those proposed
so far are simple examples.
\subsection{Solution of the properly regularised fermion-DSE}
\label{SII}
In Euclidean metric, with \mbox{$\{\gamma_\mu,\gamma_\nu\} = 2
\delta_{\mu\nu}$} and $\gamma_\mu^\dagger = \gamma_\mu$, the unrenormalised
fermion-DSE in quenched, massless, d$=$3- or 4-dimensional QED is
\begin{equation}
S^{-1}(p) = i \gamma\cdot p + e^2 \int \frac{d^{\rm d}q}{(2\pi)^{\rm d}}
        D_{\mu\nu}(p-q) \, \gamma_\mu \,S(q) \,\Gamma_\nu(q,p).
\label{DSEdQED}
\end{equation}
We use a $4\times 4$ representation of the Euclidean Dirac algebra in 3 and 4
dimensions, which allows for a study of dynamical chiral symmetry breaking
without parity violation\cite{RP84}.  For d=3 one can use, for example,
$\gamma_1$, $\gamma_2$, $\gamma_4$ with $\gamma_5 = - \gamma_1 \gamma_2
\gamma_3 \gamma_4$, where the matrices have their usual Euclidean
definitions.  Using the covariant gauge fixing procedure, as we do herein,
the fermion propagator in Eq.~(\ref{DSEdQED}) has the general form
\mbox{$ S^{-1}(p) = i\gamma\cdot p \, A(p^2) + B(p^2)$}.

The quenched photon propagator in Eq.~(\ref{DSEdQED}) is
\begin{equation}
\label{PProp}
D_{\mu\nu}(k) =
\left( \delta_{\mu\nu} - \frac{k_\mu k_\nu}{k^2}\right) \frac{1}{k^2}
        - k_{\mu}k_{\nu}\hat{\Delta}(k;\xi)
\equiv D_{\mu \nu}^{\rm T}(k) - k_{\mu}k_{\nu}\hat{\Delta}(k;\xi)
\end{equation}
where $\hat{\Delta}(k;\xi)$ is the gauge fixing term, with $\xi$ the gauge
parameter.  One particular choice is
\begin{equation}
\label{Deltap}
\hat{\Delta}(k;\xi) = - \xi\,\frac{1}{(k^2)^2}\,
R\left(\frac{k^2}{\Lambda^2}\right)~,
\end{equation}
where $\Lambda^2$ is a regularising parameter and
\begin{eqnarray}
\label{defR}
R(x)>0\;\forall x~, \;\; \mbox{with}\;\;
R(0)=1 \;& \;\;\mbox{and}\;\; & \; \int_0^\infty\,dx\, R(x) =1~.
\end{eqnarray}
Equation~(\ref{Deltap}) reduces to the standard covariant gauge fixing term
when $\Lambda^2\rightarrow\infty$.

Defining the function
\begin{equation}
\Delta(x;\xi) = \int \frac{d^{\rm d}k}{(2\pi)^{\rm d}}\,
        \hat{\Delta}(k;\xi)\,{\rm e}^{ik\cdot x}
\label{Deltax}
\end{equation}
then the fermion propagator in the gauge specified by $\xi$,
$S(x,\Delta)$, is obtained from that in Landau gauge, $\xi = 0$; i.e.,
$S(x,0)$, via the Landau-Khalatnikov-Fradkin\cite{LKF56,JZ59,BZ60} (LKF)
transformation
\begin{equation}
\label{LKFT}
S(x;\Delta) = S(x,0) {\rm e}^{e^2 [\Delta(0;\xi) - \Delta(x;\xi)]}~.
\end{equation}

The fermion--gauge-boson vertex in Eq.~(\ref{DSEdQED}) can be written in the
general form
\begin{equation}
\label{GVTX}
\Gamma_\mu(p,q) = \Gamma_\mu^{\rm BC}(p,q)
        + \sum_{i=1}^8\, T_\mu^i(p,q)\,g^i(p^2,p\cdot q,q^2)
\end{equation}
where\cite{BC80}
\begin{eqnarray}
\Gamma_\mu^{\rm BC}(p,q) & = & \Sigma_A(p,q)\,\gamma_\mu +
(p+q)_{\mu}\left\{ \Delta_A(p,q)\,\case{1}{2}\,
                \left[ \gamma\cdot p + \gamma\cdot q\right]
- i\Delta_B(p,q)\right\}~,
\label{BCV}
\end{eqnarray}
with
\begin{eqnarray}
\Sigma_A(p,q) \equiv \case{1}{2}\,[A(p^2) + A(q^2)] \; & \;\; \mbox{and}\;\; &
\; \Delta_A(p,q) \equiv \frac{A(p^2) - A(q^2)}{p^2 - q^2}~,
\end{eqnarray}
and similarly for $\Delta_B$.

In Eq.~(\ref{GVTX}), $T_\mu^i(p,q)$ are eight tensors, transverse with
respect to $(p-q)_\mu$, given in Eq.~(\ref{Ts}) of the Appendix.  Under
charge conjugation the vertex transforms as follows\cite{LS69}
\begin{equation}
\label{CCT}
\left[\Gamma_\mu(-q,-p)\right]^{\rm T} =
- {\cal C}\, \Gamma_\mu(p,q)\,{\cal C}^\dagger~,
\end{equation}
where ``T'' denotes matrix transpose and ${\cal C} = \gamma_2\gamma_4$ is the
charge conjugation matrix, which entails that all of the functions $g^i$ are
symmetric under $p\leftrightarrow q$ except for $g^6$, which is
antisymmetric.

In considering Eq.~(\ref{DSEdQED}) alone the functions $g^i$ are undetermined
although, in principle, they can be calculated within the DSE framework.  As
we have remarked, however, this is a difficult and unsolved problem.
Hitherto, progress toward determining these functions has been made by studying
the implications of the following constraints, which the fermion--gauge-boson
vertex in Abelian gauge theories must satisfy\cite{BR93}: the vertex must A)
satisfy the Ward-Takahashi identity; B) be free of kinematic singularities
[i.e., have a well defined limit as $p\rightarrow q$]; C) reduce to the
bare vertex in the free field limit in the manner prescribed by perturbation
theory; D) transform under charge conjugation as indicated in Eq.~(\ref{CCT})
and preserve the Lorentz symmetries of the theory; E) ensure local gauge
covariance of the propagators; and F) ensure multiplicative renormalisability
of the DSE in which it appears.

In general, Eq~(\ref{DSEdQED}) admits both chirally asymmetric, $B\neq0$, and
chirally symmetric, $B= 0$, solutions.  Herein we focus on the latter, in
which case the propagator has the form
\begin{equation}
\label{SCS}
S(p) = \frac{1}{i\gamma\cdot p \,A(p^2)} =
\frac{{\cal F}(p^2)}{i\gamma\cdot p }~.
\end{equation}
Only the tensors $i=2,3,6,8$ in Eq.~(\ref{GVTX}) contribute in this case and
the active part of the vertex can therefore be written as
\begin{equation}
\label{CSVTX}
\Gamma_\mu(p,q) = \left.\Gamma_\mu^{\rm BC}(p,q) \right|_{B=0}
        + \Delta_A(p,q)\,
        \sum_{i=2,3,6,8}\, T_\mu^i(p,q)\,f^i(p^2,p\cdot q,q^2)~.
\end{equation}
Substituting Eqs.~(\ref{PProp}), (\ref{SCS}) and (\ref{CSVTX}) into
Eq.~(\ref{DSEdQED}) yields
\begin{eqnarray}
\label{DSEA}
\lefteqn{S^{-1}(p)  =  i \gamma\cdot p
- \left\{e^2 \int \frac{d^{\rm d}q}{(2\pi)^{\rm d}}\,
                 i\gamma\cdot (p-q)\,\hat{\Delta}(p-q;\xi)\right\} } \\
& &
+ e^2 \int \frac{d^{\rm d}q}{(2\pi)^{\rm d}}
        \,i\gamma\cdot (p-q) \hat{\Delta}(p-q;\xi)
                 S(q)\,S^{-1}(p)
 + \,e^2 \int \frac{d^{\rm d}q}{(2\pi)^{\rm d}} D_{\mu\nu}^{\rm T}(p-q)
        \gamma_\mu S(q) \Gamma_\nu(q,p)~.
\nonumber
\end{eqnarray}

The parenthesised term in Eq.~(\ref{DSEA}) is zero in any translationally
invariant regularisation scheme.  In Ref.~\cite{CP91} a hard cutoff was used,
which violates translational invariance, leading to a spurious additional
term in the massless, chirally symmetric DSE.  This is why the solution
obtained therein is not gauge covariant.

In the quenched approximation and in the absence of dynamical chiral symmetry
breaking, a sufficient condition for gauge covariance of the fermion
propagator is that
\begin{equation}
\label{BRC}
\int \frac{d^{\rm d}q}{(2\pi)^{\rm d}} D_{\mu\nu}^{\rm T}(p-q)
        \gamma_\mu S(q) \Gamma_\nu(q,p) = 0~,
\end{equation}
for arbitrary $\hat{\Delta}(k;\xi)$\cite{BR93}.  This leads to constraints
on, and relations between, the functions $f^i$, which we discuss in
Sec.~\ref{Vf}.  With a vertex satisfying Eq.~(\ref{BRC}), Eq.~(\ref{DSEA})
reduces to the following linear equation for ${\cal F}(p^2)$:
\begin{equation}
\label{DSEGCd}
1 = {\cal F}(p^2)
+ e^2 \int \frac{d^{\rm d}q}{(2\pi)^{\rm d}}
        \,(p-q)\cdot q\, \hat{\Delta}(p-q;\xi)
                 \frac{{\cal F}(q^2)}{q^2}~.
\end{equation}
In the following we discuss the cases d$=$3 and 4.

\subsubsection{Three-dimensional QED}
The case d$=$3 was considered explicitly in Ref.~\cite{BR93}.  In this case
the regularising parameter can be removed, $\Lambda^{2}\rightarrow\infty$,
and the equation takes the form
\begin{equation}
{\cal F}_(p) = -
\frac{\alpha\xi }{2 p }\int_0^\infty \, dq\, q\,{\cal F}(q)\,
\frac{d}{dq}\left(\frac{1}{q}\ln\left|\frac{p + q}{p-q}\right|
                        \right)~,
\end{equation}
where $\alpha=e^2/(4\pi)$, and the solution is
\begin{equation}
{\cal F}(p) = 1 -
\frac{\alpha \xi}{2 p}\arctan\left(\frac{2 p}{\alpha \xi}\right)~.
\end{equation}
This is just the LKF transform of the free fermion propagator, which is as it
must be since that is the solution in Landau gauge.

\subsubsection{Four-dimensional QED}
In studying d$=$4 we consider the renormalised form of this equation:
\begin{equation}
\label{DSERFD}
1 = {\cal Z}_2 {\cal F}_R(p) +
{\cal Z}_2 e^2 \int  \frac{d^4q}{(2\pi)^4} \, (p-q).q\,
\hat{\Delta}(p-q;\xi) \frac{{\cal F}_R(q)}{q^2}~,
\end{equation}
where \mbox{${\cal Z}_2 {\cal F}_R(p) = {\cal F}(p)$}, which, because of
multiplicative renormalisability, is expected to have a power law
solution\cite{BD91}: \mbox{${\cal F}_R = (p^2/\mu^2)^{\phi(\xi)}$}, where
$\phi(\xi)$ is not determined by the constraint of multiplicative
renormalisability.  In Landau gauge \mbox{$\hat{\Delta}(x;\xi=0)=0$} and
hence from Eq.~(\ref{DSERFD}) ${\cal Z}_2^{\xi=0} {\cal F}_R(p) = 1$; i.e.,
\mbox{$\phi(\xi=0)=0$}.  Renormalising such that one has the free, massless
fermion propagator as the solution in Landau gauge then
\begin{equation}
\label{LGZ}
{\cal Z}_2^{\xi=0} = 1~.
\end{equation}

It follows from Eq.~(\ref{DSERFD}) that
\begin{equation}
{\cal F}_R(p_2) - {\cal F}_R(p_1) = e^2 \int \frac{d^4q}{(2\pi)^{\rm
d}} \frac{{\cal F}_R(q)}{q^2}
\left((p_1-q).q \,\hat{\Delta}(p_1-q;\xi)
 - (p_2-q).q\,\hat{\Delta}(p_2-q;\xi) \right)~.
\label{SubDSE}
\end{equation}
After evaluating the angular integral, the right-hand-side of
Eq.~(\ref{SubDSE}) is finite when the regularising parameter is removed,
$\Lambda^{2}\rightarrow\infty$, and this yields
\begin{equation}
{\cal F}_R(p^2) - {\cal F}_R(\mu^2) = -\frac{\alpha
\xi}{4\pi}\int_{p^2}^{\mu^2}dq^2\,\frac{{\cal F}_R(q)}{q^2}~.
\end{equation}
The solution of this equation is
\begin{equation}
\label{soln}
{\cal F}_R(p^2) = {\cal F}_R(\mu^2)
\left(\frac{p^2}{\mu^2}\right)^{\frac{\alpha\xi}{4\pi}}~,
\end{equation}
which is multiplicatively renormalisable and gauge covariant (again, this is
the LKF transform of the free fermion propagator, as it must be).

The renormalisation constant in an arbitrary gauge follows from
Eq.~(\ref{LKFT})~\cite{JZ59}
\begin{equation}
{\cal Z}_2^\xi = {\cal Z}_2^{\xi=0}\,{\rm e}^{e^2 \Delta(0;\xi)}~.
\end{equation}
Defining $H(x)= e^2 [\Delta(0;\xi)-\Delta(x;\xi)]$, one finds from
Eqs.~(\ref{Deltap}) and (\ref{Deltax}) that
\begin{equation}
x^2\,H'(x^2)  =  -\nu\,\int_0^\infty\,dk\,\frac{2}{k}\,J_2(kx)\,
        R\left(\frac{k^2}{\Lambda^2}\right)
\end{equation}
where $\nu = (\alpha\xi)/(4\pi)$ and $J_2$ is a Bessel function.

For $\Lambda$ sufficiently large the properties of $R(x)$ given in
Eq.~(\ref{defR}) entail that
\begin{equation}
x^2\,H'(x^2) \approx -\nu\,\int_0^\infty\,dk\,\frac{2}{k}\,J_2(kx)\,
        \exp\left(-\frac{k^2}{\hat{\Lambda}^2}\right)~,
\end{equation}
from which it follows that
\begin{equation}
x^2\,H'(x^2)  \approx  -\nu \frac{\hat{\Lambda}^2x^2}{1+\hat{\Lambda}^2 x^2}~,
\end{equation}
for some $\hat{\Lambda} \propto \Lambda$. Hence, introducing the scale
$\mu^2$,
\begin{equation}
e^2 \Delta(x;\xi) =  \nu\,\ln\left(\frac{\mu^2}{\Lambda^2} + \mu^2 x^2\right)~.
\end{equation}
Using Eq.~(\ref{LGZ}), this gives
\begin{equation}
{\cal Z}_2^\xi =
\left(\frac{\mu^2}{\Lambda^2}\right)^{\frac{\alpha\xi}{4\pi}}~.
\end{equation}

\subsection{Gauge covariance and vertex Ans\"atze}
\label{Vf}
\setcounter{subsubsection}{0}
The question as to the form of the vertex that satisfies Eq.~(\ref{BRC})
arises.  Substituting Eq.~(\ref{CSVTX}) into Eq.~(\ref{BRC}) leads to
\begin{eqnarray}
\label{BRCE}
\lefteqn{0  =  \int \frac{d^{\rm d}q}{(2\pi)^{\rm d}}\, \frac{1}{(p-q)^2}\,
\frac{A(p^2)-A(q^2)}{p^2-q^2}\;\times} \\
& & \left( p^2 q^2 + \case{1}{2} (p^2 + q^2)p\cdot q
                - \case{1}{2}\,p\cdot q\,\frac{(p^2-q^2)^2}{(p-q)^2}
        - f^2 (p^2 + q^2) [p^2 q^2 - (p\cdot q)^2] \right.\nonumber\\
& & \left. + 2 f^3 [ (p\cdot q)^2 + p^2 q^2 - p\cdot q\, (p^2 + q^2)]
        + f^6 (d-1) p\cdot q (p^2 -q^2)
        + f^8 (2-{\rm d}) [p^2 q^2 - (p\cdot q)^2] \rule{0mm}{6mm}\right)~.
\nonumber
\end{eqnarray}
{}From this it is clear that the vertex, $\Gamma_\mu(p,q)$, which ensures gauge
covariance and multiplicative renormalisability will, in general, depend on
$p^2$, $q^2$, $p\cdot q$ and the ratio \mbox{$\rho(p,q) = A(p)/A(q)$}.
However,
there are simple, $\rho$-independent choices for the functions $f^i$ for which
the vertex satisfies Eq.~(\ref{BRCE}).

To illustrate this we assume that the functions $f^i$ in Eq.~(\ref{CSVTX})
are independent of $p\cdot q$.  In this case one can evaluate the angular
integrals in Eq.~(\ref{BRCE}) to find
\begin{eqnarray}
\label{BRCEE}
0 & = & \int_0^\infty\,dq^{{\rm d}-2}\,{\cal F}(q^2)\,\Delta_A(p,q)\;
\left( \case{1}{2}\,({\rm d} -1)   I_3
+ ({\rm d}-1) \frac{p^2-q^2}{p^2+q^2} I_3 f^6(q^2,p^2) \right. \\
& & \left. + \left[ I_1 -  I_3\right]
        \left[f^3(q^2,p^2) - \case{1}{2}\left[p^2+q^2\right]^2 f^2(q^2,p^2)
                 + \left(1-\case{{\rm d}}{2}\right)f^8(q^2,p^2) \right]
 \right)~,
\nonumber
\end{eqnarray}
where the $I_i(p^2,q^2)$ are given in Eq.~(\ref{Is}) and we have used
Eqs.~(\ref{UR1}) and (\ref{UR2}).

\subsubsection{Curtis-Pennington Ansatz}
It is clear that the choice
\begin{eqnarray}
f^6(p^2,q^2) =  \case{1}{2}\,\frac{p^2 + q^2}{p^2 - q^2} \; & \;\;
\mbox{with}\;\;
& \; f^i = 0, \; i\neq 6
\end{eqnarray}
ensures that Eq.~(\ref{BRCEE}) is satisfied.  This is just the chirally
symmetric limit of the vertex proposed in Ref.~\cite{CP90} and so we see that
this Ansatz leads to a solution of the chirally symmetric, quenched
fermion-DSE that is gauge covariant and multiplicatively renormalisable, as
discussed in Ref.~\cite{BR93}.  [Equation (35) in Ref.~\cite{BR93} is the
equation given in Ref.~\cite{CP91}.  As we remarked above, the fact that the
parenthesised term in Eq.~(\ref{DSEA}) is zero eliminates the first term on the
right-hand-side of this equation, which becomes Eq.~(\ref{DSEGCd}) above.]

There is a problem with this vertex, however:
\begin{equation}
\lim_{p\rightarrow q} f^6(p^2,q^2)\,T_\mu^6(p,q) = \mbox{indeterminate}~.
\end{equation}
Therefore, in the chirally symmetric case, $B=0$, this vertex violates
criterion B) and hence does not satisfy the Ward identity.

\subsubsection{A Ward identity preserving Ansatz}
A simple Ansatz satisfying Eq.~(\ref{BRC}) and all of the criteria listed in
Sec.~\ref{SII}, and hence one that preserves the Ward identity in the chiral
limit, is $f^i=0$ for $i\neq 3,8$ with
\begin{eqnarray}
f^3 = \case{1}{2} \left( \case{\rm d}{2} -1 \right) f^8 \;
& \;\;\mbox{and}\;\;& \;
f^8 = \frac{1}{\case{\rm d}{2} -1} \frac{({\rm d} -1) I_3}{I_1 - I_3}~.
\end{eqnarray}

For d$=$3, $I_1$ has a logarithmic divergence as $p\rightarrow q$ [see
Eq.~(\ref{Idthree})] but nevertheless
\begin{equation}
\lim_{p\rightarrow q} f^8(p^2,q^2)\,T_\mu^8(p,q) = 0~,
\end{equation}
with a similar result for the $T_\mu^3$ term.  Hence, criterion B) is
satisfied and the vertex preserves the Ward identity.

For d$=$4 the explicit form of this vertex Ansatz is
\begin{eqnarray}
f^3 = \case{1}{2} f^8 \;&  \;\;\mbox{with}\;\; &
f^8 = \left\{
\begin{array}{ll}
\frac{3\left(p^2+q^2\right)}{3 p^2 - q^2}~,\;\; p^2>q^2\\
\frac{3\left(p^2+q^2\right)}{3 q^2 - p^2}~,\;\; p^2<q^2 \\
\end{array}~.\right.
\end{eqnarray}
It will be observed that for $p^2>\!>q^2$ one recovers the
O($\alpha$)-corrected perturbative vertex in the leading logarithm
approximation with this Ansatz~\cite{CP90}, consistent with constraint C)
above.  It is the simplest Ansatz which both does this and preserves the Ward
identity in the chirally symmetric limit, $B=0$.

\subsubsection{A more general, Ward identity preserving Ansatz}
For $d=4$, a more general Ansatz can be obtained by writing Eq.~(\ref{BRCEE})
in the following form:
\begin{equation}
\label{Explicit}
0 = \int_0^1 dx\,
\left( x [1-\rho(x)] - \frac{1}{x}\left[1 - \frac{1}{\rho(x)}\right] \right)\,
\left\{ 3 f^6(1,x) + \case{3}{2} \,\frac{1+x}{1-x} + h(1,x)\,
\frac{3-x}{1-x}\right\}
\end{equation}
with $\rho(x) \equiv A(p^2)/A(q^2)$ and
\begin{equation}
\label{hx}
h(1,x) = f^3(1,x) - \case{1}{2} (1+x) f^2(1,x) - f^8(1,x)~.
\end{equation}
In deriving Eq.~(\ref{Explicit}) we used the symmetry properties:
$f^i(1,x) = f^i(1,x^{-1})$ for $i\neq 6$ and $f^6(1,x) = - f^6(1,x^{-1})$ and
the explicit forms of the angular integrals given in the appendix.

Taking into account the symmetry properties of $f^i$ and criterion B) one can
introduce a function $F(x)$ with the properties
\begin{eqnarray}
\label{conds}
\mbox{a)}\;\;\int_0^1 dx\, F(x) = 0 \;\; & \; {\rm and} \; & \;\;
\mbox{b)}\;\; F(1) + F'(1) = 6 \rho'(1)
\end{eqnarray}
in terms of which Eq.~(\ref{Explicit}) is solved by
\begin{eqnarray}
\label{hpart}
h(1,x) & = & \case{1}{4}\frac{ F(x) - F(1/x) }
        {\displaystyle
        x [1-\rho(x)] - \frac{1}{x}\left[1 - \frac{1}{\rho(x)}\right]}~, \\
f^6(1,x) & = & -\left[ \case{1}{2} + \case{1}{3}h(1,x)\right] \frac{1+x}{1-x}
        + \case{1}{6}\frac{ F(x) + F(1/x) }
        {\displaystyle
        x [1-\rho(x)] - \frac{1}{x}\left[1 - \frac{1}{\rho(x)}\right]}~.
\end{eqnarray}
Condition b) in Eq.~(\ref{conds}) ensures that $f^6(1,1)$ is finite and hence
that the vertex is free of kinematic singularities.

In order to further constrain $F(x)$, we note that since the denominator in
Eq.~(\ref{hpart}) is of O($\alpha$) in the perturbative expansion of
$\rho(x)$ for small $\nu$, one may also require that $F(x)$ be of O($\alpha$)
in this case.  One example of a model that satisfies this constraint and
Eq.~(\ref{conds}) is
\begin{equation}
\label{formF}
F(x) = 6 \rho'(1)\,\frac{ 1- \omega \, x^{\omega - 1} }{1-\omega^2}~,
\end{equation}
with $\omega>0$ and O($\alpha$).

Equations (\ref{CSVTX}), (\ref{hx}) and (\ref{hpart}-\ref{formF}) provide a
$\rho(x)$ dependent, Ward identity preserving vertex Ansatz.

\subsection{Summary and conclusions}
We have studied the Dyson-Schwinger equation (DSE) for the fermion propagator
in quenched, massless three- and four-dimensional QED obtaining the chirally
symmetric solution in both cases.  The solutions are gauge covariant and
multiplicatively renormalisable.

The solution obtained in a previous study\cite{CP91} of the four-dimensional
case is incorrect.  We showed that the error arises because of an
inappropriate regularisation of the DSE.

We employed a constraint equation, proposed in Ref.~\cite{BR93}, to
demonstrate that gauge covariance of the solution restricts the form of the
fermion--gauge-boson vertex and showed that, in general, the vertex can
depend on the ratio $\rho(p,q)=A(p^2)/A(q^2)$.  It is not difficult to
construct
explicit examples of this type, however, simpler forms are possible.  The
Ansatz proposed in Ref.~\cite{CP90} is one such form, however, with $B=0$, it
violates the Ward identity and is therefore not suitable for studies of the
chirally symmetric fermion-DSE.  We proposed a minimal, alternate form which
overcomes this defect.  Other forms are possible and all can be constructed
using the constraint equation.

\acknowledgements
CDR acknowledges a useful correspondence with C. J. Burden.  The work of ZD
and HJM was supported in part by the US Department of Energy under grant
number DE-FG02-85-ER40214.  The work of CDR was supported by the US
Department of Energy, Nuclear Physics Division, under contract number
W-31-109-ENG-38.
\appendix
\section*{}
\hspace*{-\parindent}{\bf Vertex tensors.}

The eight tensors in Eq.~(\ref{GVTX}) are, with $k=p-q$,
\begin{eqnarray}
& & \nonumber
\begin{array}{ll}
T_\mu^1 = -i p_\mu (k\cdot q) + i q_\mu (k\cdot p)~; &
T_\mu^2 = -i \gamma\cdot (p +q) T_\mu^1~; \\
T_\mu^3 = \gamma_\mu k^2 - k_\mu \gamma\cdot k~; &
T_\mu^4 = -\case{1}{2}[\gamma\cdot p,\gamma\cdot q] T_\mu^1~;
\end{array}
\\\nonumber
& &
\begin{array}{ll}
T_\mu^5 = \case{1}{2}[\gamma_\mu, i\gamma\cdot k]~; &
T_\mu^6 = \gamma_\mu\,(p^2 - q^2) - (p+q)_\mu\,\gamma\cdot (p-q)~;
\end{array}
\\
& & \label{Ts}
T_\mu^7 =  -\case{i}{2}\,(p^2-q^2)\,[\gamma_\mu\,\gamma\cdot(p+q)
                                        - (p+q)_\mu]
                - \case{i}{2}\,(p+q)_\mu [\gamma\cdot p,\gamma\cdot q]~;
\\
& & \nonumber
T_\mu^8 = \case{1}{2}\,[ \gamma\cdot p\, \gamma\cdot q \,\gamma_\mu
                        - \gamma_\mu\, \gamma\cdot q \, \gamma\cdot p ]~.
\end{eqnarray}

\hspace*{-\parindent}{\bf Useful integrals.}

The integrals in Eq.~(\ref{BRCEE}) are
\begin{equation}
\label{Is}
\begin{array}{ll}
I_1 =  p^2 q^2 {\cal I}_1~; &
I_2 = \case{1}{2}\left(\left[p^2 + q^2\right] {\cal I}_1  - 1\right)~; \\
I_3 = \case{1}{2}\left[p^2 + q^2\right] I_2~; &
I_4 = \case{1}{4}\left[p^2-q^2\right]^2
        \left(\left[p^2 + q^2\right] {\cal I}_2  - {\cal I}_1\right)
\end{array}
\end{equation}
where
\begin{equation}
{\cal I}_n = \int\,d\Omega_{\rm d}\,\frac{1}{(p-q)^{2n}}
\end{equation}
with
\mbox{$
\int\,d\Omega_{\rm d}\, \equiv[
\frac{1}{{\cal N}} \int_0^\pi\,d\theta_{2}\,\sin^{{\rm d}-2}\theta_2
\int_0^\pi\,d\theta_{3}\,\sin^{{\rm d}-3}\theta_3 \ldots
\int_0^{2\pi}\,d\theta_{{\rm d}-1}]
$} and ${\cal N} = 2 \pi^{{\rm d}/2}/\Gamma({\rm d}/2)$. For d$=$3
\begin{eqnarray}
\label{Idthree}
{\cal I}_1  =  \frac{1}{2 p q} \ln\left(\frac{(p+q)^2}{(p-q)^2}\right)\;
& \;\; \mbox{and} \;\; & \;{\cal I}_2  =  \frac{2}{(p^2-q^2)^2}
\end{eqnarray}
while for d$=$4
\begin{eqnarray}
{\cal I}_1  =
        \frac{1}{p^2}\theta(p^2 - q^2) + \frac{1}{q^2}\theta(q^2 - p^2)\;
& \;\; \mbox{and} \;\; & \;
{\cal I}_2  =  \frac{1}{|p^2 - q^2|}\, {\cal I}_1~.
\end{eqnarray}

It can be shown that, for arbitrary d,
\begin{equation}
({\rm d} -3) \, \left(p^2 + q^2\right)\,{\cal I}_1
+ \left(p^2-q^2\right)^2\,{\cal I}_2 = {\rm d} -2
\end{equation}
from which many useful relations follow; for example,
\begin{eqnarray}
\label{UR1}
0 & = & \int\,d\Omega_{\rm d}\, \frac{1}{(p-q)^2}\,
\left( ({\rm d}-3)\, p\cdot q
+2 \, \frac{p\cdot(p-q)\,(p-q)\cdot q}{(p-q)^2}\right)~,\\
\label{UR2}
0 & = & I_1 + ({\rm d}-2 ) I_3 - I_4~.
\end{eqnarray}


\end{document}